\documentclass[aps,prb,letterpaper,showpacs,twocolumn,superscriptaddress]{revtex4}

\usepackage{color,amsmath,amssymb}
\usepackage{graphicx}
\usepackage{latexsym}
\makeatletter
\usepackage{dcolumn}
\usepackage{bm}
\usepackage{hyperref}
\usepackage{units}
\usepackage[T1]{fontenc}
\makeatletter \def\@dotsep{5} \makeatother
\makeatother

\renewcommand{\vec}{\mathbf}
\newcommand{\up}{\uparrow}
\newcommand{\dn}{\downarrow}
\newcommand\e{\mathrm{e}}
\newcommand\iu{\mathrm{i}}

\begin{document}
\title{Local Moment Formation of an Anderson Impurity on Graphene} 

\author{Chunhua Li}  
\affiliation{Department of Physics and Texas Center for Superconductivity,
University of Houston, Houston, Texas 77204, USA}
\author{Jian-Xin Zhu}
\affiliation{
Theoretical Division, Los Alamos National Laboratory, Los Alamos, New Mexico 87545, USA
}
\author{C.~S. Ting}
\affiliation{Department of Physics and Texas Center for Superconductivity,
University of Houston, Houston, Texas 77204, USA}

\begin{abstract}
We study the property of a magnetic impurity on a single-layer
graphene within an Anderson impurity model. Due to the vanishing
local density of states at the Fermi level in graphene, the
impurity spin cannot be effectively screened out.  Treating the
problem within the Gutzwiller approximation, we found a region in
the parameter space of $U$-$E^f$ where the impurity is in the
local moment state, which is characterized by a zero effective
hybridization between the bath electron and the magnetic impurity.
Here $U$ is the onsite Coulomb repulsion of the impurity electrons
and $E^f$ the impurity energy level. The competition between $U$
and $E^f$ is also discussed.  While larger $U$ reduces double
occupation and favors local moment formation, a deeper impurity
level prefers double occupation and a nonzero hybridization and
thus a Kondo screened state. For a fixed $U$, by continuously
lowering the impurity level,  the impurity first enters from a
Kondo screened state into a local moment state and then departs
from this state and re-enters into the Kondo screened state.     
\end{abstract}
\maketitle

\section{Introduction}

The research on a single layer of graphite -- graphene -- has been
advanced tremendously after its successful
exfoliation,~\cite{Novoselov2004} the importance of which cannot
be overstated. Due to its exceptional electronic, thermal, and
mechanical properties, graphene virtually finds its potential in
every aspect of modern technology. For example, it could have tremendous
applications in microelectronics due to its high electron
mobility. It also has a great deal of potential in
spintronics,~\cite{Son2006, Moghaddam2010} where the spin current
is utilized. From the theoretical perspective, it has been a test
ground for many fundamental physics laws due to its similarity to
massless Dirac fermions,~\cite{Semenoff1984, DiVincenzo1984}
where the ``speed of light'' is replaced by the graphene Fermi velocity
$v_F\sim 10^{6} \unitfrac{m}{s}$. 

The property of a magnetic impurity in a metallic host is an
important problem in condensed matter physics. In a conventional metal,
the Kondo effect, where the itinerant electrons completely screen
the magnetic moment of the impurity, is always present at low
temperatures. The impurity property is not sensitive to the
detailed band structure of the host metal. The most important element
from the impurity bath is its density of states (DOS) at the Fermi level,
$\rho(\omega = 0)$, and it enters the physics through the
scattering rate $\Gamma=\rho(0)V$, with $V$ being the
hybridization matrix between the impurity and host electrons. This is due to
the fact that in conventional metals, the DOS around the Fermi
energy is almost a finite constant. However, it has been theoretically 
found that vanishing host
DOS (power law or otherwise) could lead to significant changes to
the Kondo physics.~\cite{Withoff1990,Bulla1997} The main finding,
among many others, is that the small available states around
$\omega=0$ for conduction electrons 
makes the effective hybridization between the band
electron and the impurity vanishingly small. This ineffective
screening of the impurity moment causes the impurity to decouple
from the conduction media and form a free local moment (LM).
Withoff and Fradkin~\cite{Withoff1990} found that the Kondo coupling, $J$, has to be
larger than some critial value for the Kondo effect to take
place. However, the dependence of the magnetic impurity state on
the conduction electron DOS is a tricky one. For exmaple,
numerical renormalization group calculation shows that the
impurity state depends not only on the index of a power law DOS
but also on the size of this power law
region.~\cite{Gonzalez-Buxton1998}

The hexagonal structure of graphene has two inequivalent carbon
atoms, A and B, in one unit cell. In the tight bind approximation,
the graphene electronic structure is modelled by a nearest
neighbor hopping integral $t$. Further neighboring hopping
integrals are small and are usually neglected. The conduction and
valence bands then touch at the corners of the graphene Brillouin
zone, $\vec{K}$ and $\vec{K}'$, which are called the Dirac points.
The dispersion near these Dirac points is linear in $\vec{k}$ and
the DOS is also linear in energy. For all practical purposes, this
tight binding description is quite accurate.
Many peculiar
properties of graphene are closely related to this linear
dispersion and the corresponding linear DOS at low energies.~\cite{AHCastroNeto:2009} 
 The
state of a magnetic impurity on graphene is expected to be very
different from that on an ordinary metal. However, the existence
of two inequivalent Dirac points could also complicate the simple
linear DOS argument since this may lead to multi-channel Kondo
effect.~\cite{Zhu2010j} The specific problem of a magnetic
impurity on graphene has recently been studied by several
groups.~\cite{Uchoa2008, Zhuang2009, Cornaglia2009, Jacob2010, MVojta2010}
Besides its theoretical importance as a prototypical many-body toy
model, it is also important to understand magnetic impurities on
graphene from the application perspective.   It was observed that
introduction of vacancies into graphite or graphene could indeed
induce ferromagnetism.~\cite{Esquinazi2003, Ohldag2007,
Barzola-Quiquia2007, Chen2010g}

In this paper, we explore the properties of a magnetic impurity in a single-layer 
graphene within an Anderson impurity model. Within the Gutzwiller approximation, 
we are able to obtain a phase diagram within the parameter space 
spanned by the impurity level and on-site Hubbard repulsion. We identify a parameter region, 
where local moment is formed on the impurity. In addition, a re-entrant Kondo resonance behavior is found. The effect of graphene chemical
potential, $\mu$, is also discussed.

The outline of the paper is as follows: In Sec.~\ref{SEC:Model}, the model Hamiltonian 
for an Anderson impurity in the graphene is introduced. The Gutzwiller procedure toward the solution is explained. In Sec.~\ref{SEC:Results}, we present the results on the local moment 
formation and Kondo resonance state as a result of the competition between the impurity level 
and on-site Hubbard repulsion. Finally, a summary is given in Sec.~\ref{SEC:Summary}.

\section{Model Hamiltonian and Theoretical Method}
\label{SEC:Model}
We start with  a
single-impurity Anderson model in graphene. The Hamiltonian is partitioned
into three parts---  the pristine graphene subsystem $\hat{H}_0$, the impurity subsystem
 $\hat{H}_{\rm imp}$, 
and the hybridization $\hat{H}_{\rm hyb}$ between these two subsystems, that is, 
\begin{equation}
    \mathcal{H} = \hat{H}_0 + \hat{H}_{\rm
    imp} + \hat{H}_{\rm hyb} \;, \label{eq:Ham}
\end{equation}
where the three parts can be written out explicitly,
\begin{subequations}
    \begin{align} 
    \hat{H}_0 =& -t\sum_{\vec{k}\sigma}\xi(\vec{k})
    a_{\vec{k}\sigma}^\dagger b_{\vec{k}\sigma} + \mathrm{H.c.}\;,
    \\ H_{\rm
    imp} =& \sum_\sigma E^f f_\sigma^\dagger f_\sigma + U
    f_{\up}^\dagger f_\up f_\dn^\dagger f_\dn\;, \\
    H_{\rm hyb} =& \frac{V}{\sqrt{\mathcal N}} \sum_{\vec{k}}
    a_{\vec{k}\sigma}^\dagger f_\sigma + \mathrm{H.c.}\;.
\end{align}
\end{subequations}
We take a $\vec{k}$-independent hybridization between carbon
$\pi$-electrons and the impurity spin, $V_\vec{k} = V$.  Here
$\xi(\vec{k}) = \sum_{i=1}^3\e^{\iu\vec{k}\cdot \vec{\eta}_i}$,
where $\vec{\eta}_i$, $i = 1, 2, 3$, are the vectors connecting to
the three nearest neighbors of a carbon site, and $\mathcal{N}$ is
the number of unit cells. As mentioned above, $t$ is the nearest
neighbor hopping integral, $t=2.7$eV. In the rest of the paper we
take $t$ as the unit of energy. The impurity is taken to be at the
origin of the A-sublattice. For the more general case where a
magnetic adatom is considered, the impurity could be anywhere and
hybridization between the impurity and its all neighboring sites
should be considered. As discussed by Uchoa {\em et 
al.},~\cite{Uchoa2011} this leads to an effective
$\vec{k}$-dependent hybridization $V(\vec{k})$.

The rich physics of the Anderson impurity model comes from the
on-site Hubbard interaction $U$ on the impurity. For small $U$, a Hartree-Fock
approximation is usually sufficient. For the cases of  intermediate to large $U$,
more degrees  of the correlation effect must be taken into
account. Here we take a projection approach and treat $U$ in the
Gutzwiller approximation.  In this approximation, the effect of
$U$ is taken into account by reducing the effective hybridization
between the conduction electrons and the impurity electrons. For
example, in the $U=\infty$ limit, the process $f_\sigma^\dagger
a_{0\sigma}$ is prohibited if there is already a $\bar\sigma$-$f$
electron at the impurity site, while there is no such restriction
if $U=0$; see Ref~\cite{Zhang1988} for a detailed discussion. On the mean
field level, we renormalize the hybridization strength $V$ with
$\widetilde{V}_\sigma=g_\sigma V$, $g_\sigma \le 1$, and replace the interaction term
with $Ud$, with $d= \langle f_\up^\dagger f_\up f_\dn^\dagger
f_\dn\rangle$. Here $g_\sigma$ is the Gutzwiller factor that depends
on the impurity number occupation $n^f_\sigma$
($n^f=n^f_\up+n^f_\dn$ being the total impurity occupation) and
double occupation $d$. The spin dependent Gutzwiller factor is,
\begin{equation}
    g_\sigma = \frac{\sqrt{(1-n^f+d)(n^f_\sigma-d)} +
    \sqrt{d(n^f_{\bar{\sigma}}-d)}}
    {\sqrt{n^f_\sigma(1-n^f_\sigma)}}. \label{eq:gsigma}
\end{equation}
Note that this form of $g_\sigma$ explicitly breaks the spin
rotational invariance, which the model Hamiltonian possesses.
However, in the following, we will only concern ourselves with the
stability of the nonmagnetic Kondo screened state, $n_\up=n_\dn$,
and thus concentrate on the calculation where $g_\up=g_\dn=g$.  A
constraint term, $\sum_\sigma\lambda_\sigma(\hat{n}_\sigma^f -
n_\sigma^f)$, is also added to the mean field Hamiltonian.  Here
the Langrange multipliers, $\lambda_\sigma$, serve to renormalize
the impurity energy level. Although this term is formally the same
to a direct Hartree decoupling of the interaction term,
$\sum_\sigma U\hat{n}_\sigma^fn_{\bar{\sigma}}^f$, the
self-consistency conditions are quite different, and thus would
lead to different physics.  It is noted that this approach is
equivalent to the Kotliar-Ruckenstein slave-boson
approximation,~\cite{Kotliar1986} 
which has been successfully applied
in the study of Anderson impurity models.~\cite{Savrasov2005} After the
above mean field procedure, we have the renormalized hybridization
and impurity Hamiltonians,
\begin{align}
    &\widetilde{H}_{\rm hyb}= \frac{1}{\sqrt{\mathcal N}}
    \sum_{\vec{k}\sigma} \widetilde{V}_\sigma
    a_{\vec{k}\sigma}^\dagger f_\sigma + \mathrm{H.c.},\\
    &\widetilde{H}_{\rm
    imp}=\sum_\sigma\left(E^f+\lambda_\sigma\right)f_\sigma^\dagger
    f_\sigma +Ud.
\end{align}
As such, the effective non-interacting Hamiltonian is,
$\widetilde{\mathcal H} = H_0 + \widetilde{H}_{\rm hyb} +
\widetilde{H}_{\rm imp}$. There are two major differences
comparred to the Hartree-Fock mean field approximation. First,
the hybridization strength $V$ is renormalized with a factor $g_\sigma$
that is dependent on the impurity occupation $n^f_\sigma$ and
double occupation $d$, which is in turn dependent on $U$. The
second difference is that the impurity level is renormalized by
the Lagrange multipliers $\lambda_\sigma$, instead of
$Un_{\bar{\sigma}}^f$. In the Hartree-Fock approximation, the impurity
level renormalization $Un_{\bar{\sigma}}^f$ becomes unphysically
large for large $U$.

In the limit $U \rightarrow \infty$, the double occupation $d$
should tend to zero. By looking at the form of $g_\sigma$ in
Eq.~(\ref{eq:gsigma}), it is clear that if at the same time $n^f$
goes to 1, then $g_\sigma$ approaches zero. In this limit, the
magnetic impurity and the conduction band decouples since the
effective hybridization $\widetilde{V}_\sigma=g_\sigma V$ is now
zero. The impurity is in the LM state. It is interesting to
realize that this is analogous to the Brinkman-Rice type of 
metal-insulator transition in the Hubbard model at half
filling,~\cite{Brinkman1970} except that here one requires ``half
filling'' on the impurity site. Unlike the half filled Hubbard
model, in addition to the chemical potential $\mu$, the impurity
number occupation is also dependent on several other aspects. It
obviously is a function of the impurity level $E^f$. It also
depends on $U$, since the interaction modifies the impurity 
level through $\lambda_\sigma$. The conduction electronic
structure and the hybridization $V$ also affect the impurity
occupation. One can then tune these parameters so that  the
impurity site is half filled,
and then expects to have an LM state for appropriate
values of $U$, where $d=0$. It is also worth pointing out that
having a state with $n^f=1$ does not mean that the impurity
 is in the LM state.
The criterion is to have a zero $g_\sigma$. This will become
apparent when we present the results below.

\begin{figure}[htb]
    \centering
    \includegraphics[clip,angle=0,origin=c,width=0.4\textwidth]
    {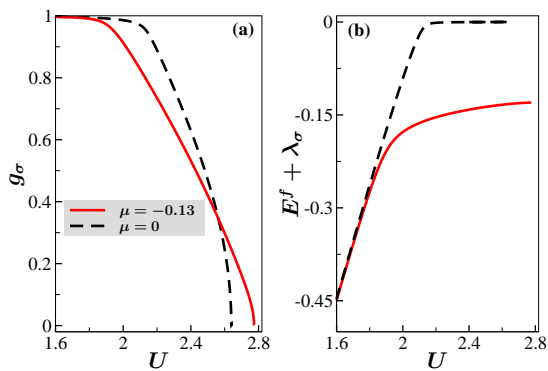} \caption{(Color online) The Gutzwiller factor $g_\sigma$ (a) and the
    effective impurity level $E^f+\lambda_\sigma$ (b) as a
    function of the on-site interaction $U$ for two different
    chemical potential values $\mu = 0$ and $\mu = -0.13$. The
    impurity level is fixed at $E^f = -2$ and the hybridization
    strength is set at $V=0.52$.  }
    \label{fig:guU}
\end{figure}

We now set out to solve the effective Hamitonian
$\widetilde{\mathcal H}$. First of all, $\widetilde{\mathcal H}$
is now non-interacting and can be readily solved by the usual
Green function method, except that we need to take care of the
self-consistency equations of $d$ and $\lambda_\sigma$, which are
obtained by minimizing the free energy. At zero temperature,
$T=0$ (on which the rest of the calculation is based), the free
energy equals to the ground state energy and the
self-consistency equations are obtained by taking partial
derivatives of $\langle \widetilde{H}\rangle $ with respect to $d$
and $n^f_\sigma$:
\begin{align}
    & U + V\sum_{\sigma} \frac{\partial g_{\sigma}} {\partial d}
    \left[\left\langle a_{0\sigma}^\dagger f_{\sigma}
    \right\rangle + \left\langle f_{\sigma}^\dagger
    a_{0\sigma}\right\rangle\right] = 0, \label{eq:d:self}\\ &
    \lambda_\sigma = V\sum_\sigma\frac{\partial g_{\sigma}}
    {\partial n^f_\sigma} \left[\left\langle a_{0\sigma}^\dagger
    f_{\sigma} \right\rangle + \left\langle f_{\sigma}^\dagger
    a_{0\sigma}\right\rangle\right], \label{eq:lambda:self}
\end{align}
where the expectation values $\langle\cdots\rangle$ are taken in
the effective Hamiltonian $\widetilde{\mathcal H}$ and are determined
from their respective retarded Green functions. Applying the
equation of motion method, we find the following retarded Green functions,
\begin{subequations}
\begin{align}
    G^{\mathrm R}_{f_\sigma,f_\sigma^\dagger}(\omega) &= \frac{1}{
    \omega^+ - E^f - \lambda_\sigma - \widetilde{V}_\sigma^2
    G_{0}^\mathrm{R}(\omega)},\label{eq:Gff}\\ G^{\mathrm
    R}_{a_{\vec{k}\sigma},f_\sigma^\dagger}(\omega) &=
    \frac{\omega^+} {(\omega^+)^2 - \lvert\xi(\vec{k})\rvert^2}
    \widetilde{V}_\sigma
    G^\mathrm{R}_{f_\sigma,f_\sigma^\dagger}(\omega),
\end{align}
\end{subequations}
where $\omega^+=\omega+\iu 0^+$ and $G_{0}^\mathrm{R}(\omega)$ is
the retarded Green function for  the clean graphene:
\begin{align}
    G_{0}^\mathrm{R}(\omega) = \frac{1}{\mathcal N}\sum_{\vec{k}}\frac{\omega^+}
    {\left( \omega^+ \right)^2 -\lvert \xi(\vec{k})\rvert^2}.
\end{align}
The expectation values in Eqs.~(\ref{eq:d:self}) and
(\ref{eq:lambda:self}) and $n_\sigma^f=\langle f_\sigma^\dagger
f_\sigma\rangle$ are then computed as $\langle O_1 O_2 \rangle =
-\frac{1}{\pi}\int d\omega\; \mathrm{Im}
G_{O_1O_2}^\mathrm{R}(\omega)f_{FD}(\omega-\mu)$, with $f_{FD}$
the Fermi-Dirac distribution function. For a given set of
parameters, \{$U$, $E^f$, $V$, $\mu$\}, Eqs.~(\ref{eq:d:self})
and (\ref{eq:lambda:self}) together with the form of
$\widetilde{H}$ yield a set of nonlinear equations of $d$,
$n^f_\sigma$ and $\lambda_\sigma$, which can be readily solved.
The impurity DOS is given by $\rho^f_\sigma(\omega)=-\frac{1}{\pi}
\mathrm{Im}\; G_{f_\sigma, f_\sigma^\dagger}^\mathrm{R}(\omega)$.

\section{Results and Discussions}
\label{SEC:Results}
\begin{figure}[htb]
    \centering
    \includegraphics[clip,angle=0,origin=c,width=0.45\textwidth]
    {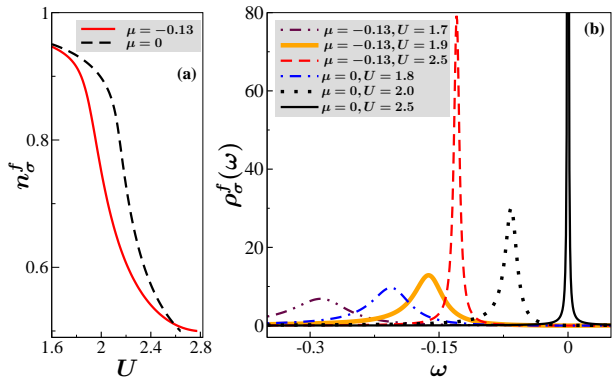} \caption{(Color online) (a) The impurity occupation as a function of
    the onsite interaction $U$ for $\mu = 0$ and $\mu = -0.13$. In
    both cases, the spin-resolved occupation number goes to $1/2$
    as $U$ approaches the critical value. (b) The resonance peak
    around $\mu$ for several values of $U$. The resonance peak
    moves progressively toward $\mu$ and the peak becomes sharper
    and more symmetric as $U$ is increased.  The impurity level is
    fixed at $E^f = -2$ and the hybridization strength is set at
    $V=0.52$.  }
    \label{fig:nres}
\end{figure}

Let us first fix the impurity level $E^f$ and see the effect of
$U$ on the impurity state.  Figure~\ref{fig:guU}(a) shows the
Gutzwiller factor as a function of $U$ for two different cases:
$\mu = 0$ and $\mu = -0.13$. The impurity level is at $E^f = -2$.
For small $U$ the Gutzwiller factor $g_\sigma$ is unity, which is
expected, since in the non-interacting impurity situation, there
should be no renormalization of the hybridization strength. As $U$
increases, $g_\sigma$ continuously decreases. There is a critical
value of $U$, $U_c$, in each case where $g_\sigma$ is zero. At
this point, the effective hybridization between the conduction
electron and magnetic impurity is zero and an LM state forms.  As
discussed earlier, the impurity occupation has to be $n_\sigma^f =
1/2$ in the LM state (see Fig.~\ref{fig:nres}(a)).  The formation
of the impurity LM state can be envisioned from two facts:
$g_\sigma$  approaches zero and the effective impurity level
$(E^f+\lambda_\sigma)\rightarrow \mu$ as $U$ goes toward 
$U_c$ (see Fig.~\ref{fig:guU}(b)). From the Green function
Eq.~(\ref{eq:Gff}), it is then clear that, if $\widetilde{V}_\sigma
\rightarrow 0$ at the same time, the impurity DOS is narrowed to a
level just at the Fermi energy. That is, the decoupling from the
conduction band and the effective impurity level reaching the Fermi
energy have to happen at the same time for an LM state to be
realized. The LM state is stable against further increase of $U$
since the impurity is already decoupled from the conduction band.
Larger $U$ will not change the double occupation, which is already
at its lowest possible value $d=0$, and thus will not change the
impurity occupation as can be seen from
Eq.~(\ref{eq:lambda:self}).  Note that before the system reaches
the LM state, the impurity occupation does depend on $U$ by
shifting the effective impurity level through $\lambda_\sigma$ and
in this regime $d$ is still dependent on $U$.

The different critical value $U_c$ for the two cases, $\mu = 0$
and $\mu = -0.13$ suggests that graphene DOS at the Fermi level
plays a central role. The graphene DOS is zero at $\mu = 0$, and
the hybridization between the host and impurity is inefficient to
fully screen out the impurity spin. By comparison, for the $\mu=-0.13$
case, where there is a finite DOS at the Fermi level, a larger
value of $U_c$ is needed to bring the impurity into the LM state. 
We can look at how the resonance peak evolve as we ramp up $U$. As
shown in Fig.~\ref{fig:nres}(b), the resonance peak moves toward
the Fermi level and also gets sharper for increasing $U$. Due to
the singularity of the graphene DOS at $\omega=\pm t$, the
impurity DOS also has peaks around these energies (not shown in
figure).  But as $U$
increases, the weight of these states diminishes and the impurity
occupation mainly comes from around the resonance energy. As the
resonance peaks move from below toward the Fermi energy, they also
become more symmetric, suggesting the effect of
$\widetilde{V}^2_\sigma G_{00}(\omega)$ term in Eq.~(\ref{eq:Gff})
falls down. 

\begin{figure}[htb]
    \centering
    \includegraphics[clip,angle=0,origin=c,width=0.4\textwidth]
    {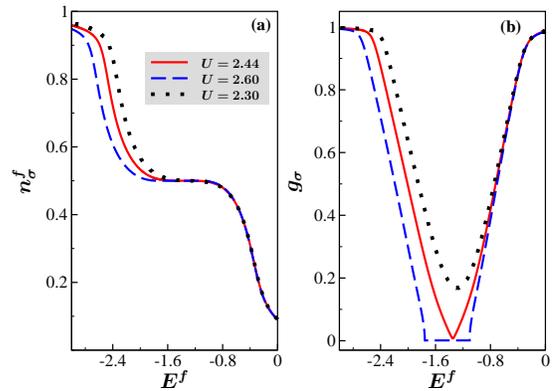} \caption{(Color online) (a) Impurity occupation as a function of $E^f$
    for three different values of $U$. (b) The Gutzwiller factor
    $g_\sigma$ as a function of $E^f$ for the same three $U$'s as
    in (a); $V=0.52$, $\mu = -0.13$ in both plots.}
    \label{fig:ng}
\end{figure}

Let us now look at the effect of the impurity level.  Unlike the
case of varying $U$, the LM state is not stable if one lowers the
impurity level past a critical value.  Figure~\ref{fig:ng}(a)
shows the impurity occupation as a function of the impurity level
for three different values of $U$. We can see that as the impurity
level goes deeper inside the conduction Fermi sea, the occupation
increases. It is interesting to notice that there is a flat region of
the curve,  where the impurity occupation seems independent of $E^f$
for moderate on-site Hubbard interaction $U$, for example $U=2.6$ in
Fig.~\ref{fig:ng}(a).  The impurity occupation in this region is
1, which suggests that an LM state has possibly formed. For
smaller $U$, the flat region shrinks and is slightly dependent on
$E^f$, for example, $U=2.3$ in Fig.~\ref{fig:ng}(a).  However,
although the curves in Fig.~\ref{fig:ng}(a) look similar, the
physics are fundamentally different for these different values of
$U$. The large $U$ value curve represents a transition from a
Kondo screened state to an LM state and then back to a Kondo state
as the impurity level is further lowered, while there is no such
transition for the smaller $U$ curve.  This can be more
convincingly seen from Fig.~\ref{fig:ng}(b) where $g_\sigma$ is
plotted against $E^f$ for the same values of $U$ as in
Fig.~\ref{fig:ng}(a). For $U=2.6$, the Gutzwiller factor
$g_\sigma$ decreases as $E^f$ goes down, reaching zero at the
upper critical value $E^f_{\rm c1} = -1.1$.  Before $E^f$ reaches
the lower critical value $E^f_{\rm c2}=-1.76$, the only solution
to the system is the decoupled conduction band and impurity. The
region between $E^f_{\rm c1}$ and $E^f_{\rm c2}$ also corresponds
to the flat region where the impurity occupation is 1 in
Fig.~\ref{fig:ng}(a). The upper critical value $E^f_{\rm{c}1}$
and lower critical value $E^f_{\rm{c}2}$  of $E^f$ are identified
from the zero segment in $g_\sigma$, which is much clearer than
from the flat region of impurity occupation as in
Fig.~\ref{fig:ng}(a).  One can also see from
Fig.~\ref{fig:ng}(b) that smaller $U$ does not produce an LM
state for all valules of $E^f$: The Gutzwiller factor is always
finite for $U=2.3$. There is a critical value of $U$ such that
below this value there is no LM state for any value of $E^f$. This
is identified as when $E^f_{\rm{c}1} = E^f_{\rm{c}2}=-1.35$ for
$U=2.44$. Comparing Fig.~\ref{fig:ng}(a) and
Fig.~\ref{fig:ng}(b), we see that $g_\sigma = 0$ is a more
natural and physically clear way of identifying the LM state.

\begin{figure}[htb]
    \centering
    \includegraphics[clip,angle=0,origin=c,width=0.4\textwidth]
    {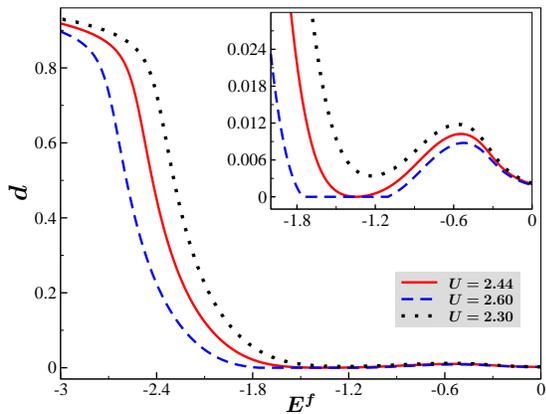} \caption{(Color online) Impurity double occupation $d$ as a function of
    the impurity level for the same three values of $U$ as in
    Fig.~\ref{fig:ng}. Inset: Zoom in of the main figure for
    $E^f$ from -2 to 0. }
    \label{fig:d}
\end{figure}

The entrance and departure of the LM state upon lowering the
impurity level suggests the competition between $U$ and $E^f$.
Large $U$ would inevitably put the impurity in the
no-double-occupancy state, $d = 0$.  In Fig.~\ref{fig:d} we
plot the double occupation, $d$, against $E^f$. When $E^f$ is
close to zero, due to small available graphene electron DOS, the
impurity occupation is small and thus the double occupation is
also small, since $d < n^f_\sigma$.  Lowering $E^f$ will increase
$d$ by increasing the overall impurity occupation $n^f$. But
eventually the effect of $U$ kicks in and further lowering $E^f$
actually makes double occupation unfavorable.  As $n^f$ approaches
1 and $d$ goes to 0, $g_\sigma$ approaches 0, where we have an LM
state. But further lowering $E^f$ favors a state, where $n^f$ is
larger than  1.  This inevitably allows double occupation and
moves the impurity out of the LM state and back to the Kondo
screened state again.

\begin{figure}[htb]
    \centering
    \includegraphics[clip,angle=0,origin=c,width=0.4\textwidth]
    {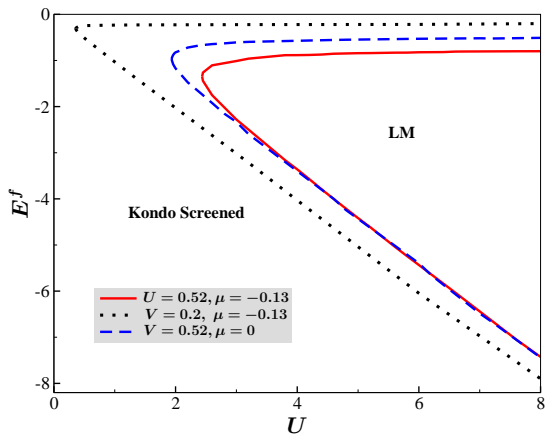} \caption{(Color online) The regions of LM state and Kondo screened state
    on the $U$-$E^f$ plane for different sets of $V$ and $\mu$.}
    \label{fig:EU}
\end{figure}

Finally, we put together the effects of $U$ and $E^f$ on the
impurity state and map out the regions of Kondo screened state and
LM state on the $U$-$E^f$ plane. We plot the ``phase diagram'' of
the impurity state in Fig.~\ref{fig:EU} for three sets of $(\mu,
V)$.  The general feature is that for moderate $U$, as we lower
the impurity level, the impurity goes from the Kondo screened
state to a decoupled LM state, and stays in this state until the
level is low enough. At this point, the impurity and the host
graphene start to talk to each other again.  However, if we look
along the $U$-axis, after the impurity enters the LM state, it
remains in that state even if we further increase $U$.  Comparing
the sets of curves, we see that larger hybridization strength
favors the Kondo screened state and thus gives a smaller LM
region. Also, moving away from the Dirac points makes it less
susceptible to LM formation.

\section{Summary and Concluding remarks} \label{SEC:Summary}
In conclusion, based on the Gutzwiller approximation, we have
carried out a detailed study on the LM formation of a single
magnetic impurity in graphene. Due to its linear DOS near the
Fermi level, graphene allows the possibility of LM formation. It
has been shown that for moderate impurity level  and large on-site
Hubbard interaction on the impurity, an LM state is favored over a
Kondo singlet state. We have explored the competition of impurity
level and on-site Hubbard interaction and mapped out the
LM region on the $U$-$E^f$ plane. We emphasize the importance of
identifying the LM state by the vanishing of the effective
hybridization as opposed to the unit impurity occupation, $n^f=1$. The effect
of graphene Fermi level is also discussed. It is remarkable that
the simple Gutzwiller approximation captures the essential physics
in this model. The Gutzwiller approximation being a mean field
approximation of the projection method does not handle the excited
states very well. But as we have seen that the spetral weight of
the impurity state is predominantly centered around the resonance
energy. This feature justifies the applicability of the method
used in this study. 

The results presented here for a single Anderson impurity also
remind us of the so called Kondo breakdown (KB)
phenomena~\cite{Pepin2007, Paul2007} in a lattice model, where a
heavy $f$ band decouples from a wider $c$ band. In the KB regime, the
decoupled $f$ band would usually be polarized due to exchange
interaction among the $f$-electrons.  The magnetic state of the LM
impurity is not determined in the single impurity case. A small
magnetic field will align it in the direction of the field, due to
the Zeeman energy. For more than two impurities, RKKY interaction
occurs, and the magnetic state of the impurities in the LM state
depends on the nature of this interaction. For example, for two
impurities situated at the lattice sites the RKKY interaction
could be ferromagnetic or antiferromagnetic, depending on whether
they are on the same or opposite sublattice.~\cite{Brey2007,
Saremi2007, Rappoport2009} The sign and strength of the RKKY
interaction also depend on the graphene doping level, and whether
the impurities are away from the carbon atoms. In such
circumstances, the real-space Gutzwiller approximation is an ideal
tool for investigating the impurity magnetic states.

\begin{acknowledgments}
We thank Dr.\ Y.\ Gao for helpful
discussions.  This work was supported by the Texas Center for
Superconductivity and the Robert A. Welch Foundation under grant
number E-1146 (C.L. \& C.S.T.), and 
by U.S. DOE  at
LANL  under Contract No. DE-AC52-06NA25396 and the DOE Office of Basic Energy
of Sciences (J.-X.Z.).
\end{acknowledgments}

\end{document}